\documentclass[conference]{IEEEtran}
\IEEEoverridecommandlockouts

\usepackage{graphicx} 
\usepackage{xcolor}
\usepackage{cite}
\usepackage{amsmath,amssymb,amsfonts}
\usepackage{algorithmic}
\usepackage{textcomp}
\usepackage{xcolor}
\usepackage{listings}
\usepackage{longtable}
\usepackage{balance}
\usepackage{longtable}
\usepackage{url}
\usepackage{booktabs} 
\usepackage{caption}  
\usepackage{siunitx}  

\def\BibTeX{{\rm B\kern-.05em{\sc i\kern-.025em b}\kern-.08em
    T\kern-.1667em\lower.7ex\hbox{E}\kern-.125emX}}

\begin{document}

\title{Taming the Security-Energy Paradox: A Green AI Approach to Optimized Android Malware Detection}
\author{
\IEEEauthorblockN{Shrinidhi Sridhar}
\IEEEauthorblockA{
\textit{MIE-SPPU Institute of Higher Education}\\
Doha, Qatar
}
\and
\IEEEauthorblockN{Vikas K. Malviya}
\IEEEauthorblockA{
\textit{MIE-SPPU Institute of Higher Education}\\
Doha, Qatar
}
}


\maketitle
\thispagestyle{plain}
\pagestyle{plain}

\begin{abstract}
An increase in advanced Android malware requires the use of deep learning models, which can run on Android devices. But there is a trade-off between security and energy use, as strong detection models can drain the battery of devices fast. This work tests different Multi-Layer Perceptron (MLP) model configurations to balance malware detection performance and energy efficiency. In this work, we compared standard FP32 models with optimized INT8 quantized neural networks with different model depths using TUANDROMD and DREBIN datasets for both classification performance and energy consumption.
The results show that INT8 quantization reduces model size by about 3.5 times with a decrease in energy consumption to 0.0189 mJ per inference, while maintaining more than 99.2\% detection accuracy. We found that shallow quantized architectures, such as 3-layer and 4-layer QNNs, reduce energy costs by improving throughput and shortening the time of CPU operating in a high-power state. This work shows that efficient malware protection can be achieved on resource-constrained smartphones and provides a foundation for Green AI in mobile security.


\end{abstract}


\section{Introduction}
The Android ecosystem has become the leading global mobile platform due to rapid growth in app distribution and a large number of active users \cite{statcounterMobileOperating,S_R19, S_R105}. This growth has given rise to more sophisticated security threats. Currently, thousands of applications are released monthly through the Google Play Store alone. This necessitates a scalable and automated screening approach to detect and manage potentially harmful applications \cite{S_R19}. At the same time, malware has also evolved from simple signature-based evasion to include polymorphic, metamorphic, and zero-day variants. This presents challenges to the traditional security frameworks \cite{S_R25, S_R46}.

Machine learning (ML) and deep learning (DL) are being used in modern mobile security to counter these threats. They offer high accuracy detection by analyzing static permissions and dynamic behavioral patterns \cite{S_R25, S_R19}. These models perform very well in controlled environments, but deploying them on physical mobile devices creates a technical challenge of high energy consumption \cite{S_R20, S_R61}. Smartphones and portable devices have limited battery capacity. Studies show that users are less likely to accept security software if it reduces battery life \cite{S_R195}. 

This issue gives rise to the accuracy-sustainability trade-off in mobile cybersecurity. Traditional Android malware detection research focuses on maximizing accuracy, precision, and recall, and neglects the environmental and operational costs of the computations \cite{S_R61, S_R97}. Complex neural network architectures are accurate, but they demand higher computational resources and energy. Sometimes this demand goes up to fifty times more than simpler models
\cite{S_R178}. Since security scans run as a continuous background process, these resource-heavy models can drain the battery quickly, due to which users may have to disable important protections \cite{S_R102}.

Green AI provides ways to address these inefficiencies by balancing resource usage with predictive performance \cite{S_R37, S_R61}. Techniques like model pruning and quantization (e.g., INT8) are being utilized as strategies to adapt high-performing models for resource-constrained edge devices \cite{S_R66, S_R173}. These techniques lower the energy consumption by reducing the precision of weights and activations. This results in lower energy consumption on the device with minimal impact on security \cite{S_R210}.

This work examines how to balance security and power consumption in Android devices. We analyzed the effects of post-training quantization and depth of different malware detection models on detection efficiency and energy consumption. From this analysis, we identified the optimal configuration to achieve optimal efficiency with less power consumption. By identifying the configurations that balance strong security with low power consumption, this work provides a scalable framework for Android security.   


The rest of the paper is structured as follows. Section \ref{sec:background} provides background on Android architecture, Android security architecture, power management, resource constraints, and the trade-off between security and energy consumption. Section \ref{sec:related} reviews existing literature related to the energy consumption in malware detection. Section \ref{sec:research_methodology} describes our approach. Section \ref{sec:results_n_discussion} presents the results of the evaluation of the approach and discusses the outcomes. 
Section \ref{sec:conclusion} concludes the paper with future work.


\section{Background} \label{sec:background}

This section provides the technical foundation of the Android ecosystem, its multi-layered security architecture, and the mechanisms used for energy consumption. It also explores the conflict between strong security protocols and the limited power of mobile devices.

\subsection{Android Architectural Framework}

Android is an open-source software stack built for mobile devices centered on a Linux kernel and is maintained by the Android Open Source Project. The Android architecture consists of multiple layers, such as the Linux kernel, native libraries, hardware abstraction layer (HAL), Android Runtime (ART), system services, and the application framework. The Linux Kernel forms the base layer of the AOSP architecture, which interacts with the hardware components. It provides core functionalities such as memory management, process management, networking, and security settings.  Above the kernel, the Hardware Abstraction Layer (HAL) provides standard interfaces for hardware-specific components (e.g., camera, sensors). Modern Android versions utilize the Android Runtime (ART), which employs Ahead-of-Time (AOT) compilation. From an energy perspective, ART has an important role in optimization. It reduces CPU cycles and instruction cache misuse by compiling app bytecode into machine code during installation rather than at runtime, which preserves battery life. \cite{androidPlatformArchitecture}.

\subsection{Security Architecture and Sandbox Model}
Android's security philosophy is based on the principle of least privilege. This is enforced through many core mechanisms:

\begin{itemize}
    \item \textbf{Application Sandboxing:} Android system utilizes Linux-based User IDs (UIDs) to isolate applications. Each application works within its own process and file system space. This prevents unauthorized access to other applications' data without cryptographically signed permissions \cite{53044}. 
    \item \textbf{SELinux (Security-Enhanced Linux):} Android uses Mandatory Access Control (MAC) through SELinux. SELinux enforces a system-wide policy that restricts process interactions even if a process is running with administrative (root) privileges. It reduces the risk of privilege escalation \cite{53044}.
    \item \textbf{Verified Boot (dm-verity):} Android uses a hardware-protected root of trust to ensure the integrity of the system image. During the boot sequence, each stage verifies the cryptographic signature of the next stage. This process prevents the persistent execution of rootkits or tampered firmware \cite{53044}
\end{itemize}

\subsection{Power Management and Resource Constraints}
Power consumption is a critical constraint in mobile computing. Android manages this using several system-level strategies which are designed to maximize deep sleep states \cite{androidPowerManagement}:

\begin{itemize}
    \item \textbf{Doze Mode and App Standby:} These features delay background network activity and CPU tasks when the device is idle. The system combines these tasks into maintenance windows. Due to this, the frequency of radio and SoC (System-on-Chip) wake-ups gets reduced.
    \item \textbf{Wake Locks:} Wake locks are used by developers to prevent the device from entering a low-power state. Wake locks are essential for functions like GPS navigation or active security scanning. But poorly optimized wake locks, which are called No-Sleep Energy Bugs, can rapidly deplete the battery.
    \item \textbf{Battery Saver Mode:} In battery saver mode, the system reduces CPU frequency, restricts background data usage, and lowers the brightness of the display to extend the battery life.
\end{itemize}

\subsection{The Interplay of Security and Energy Consumption}
Security–energy trade-off is a big challenge in Android security research. Stronger security increases computational demands. Features such as real-time threat detection, secure authentication, and continuous monitoring increase processing overhead and energy use.

Energy \textit{E} required for a secure operation can be calculated as per the following equation \cite{HU2020681}:

\begin{equation} \label{eq_energy_calculation}
	E = P_{cpu} \times t_{enc} + E_{io}
\end{equation}

Where $P_{cpu}$ is the power consumed by the processor, $t_{enc}$ is the time taken for encryption/decryption, and $E_{io}$ is the energy for I/O operations. As security protocols become more complex, $t_{enc}$ increases, resulting in a constant rise in energy consumption.

\subsection{Projects Treble and Mainline}
Systemic fragmentation has been the biggest challenge to Android security. The tight coupling between the OS framework and hardware-specific drivers delayed the delivery of critical security patches. Google introduced Project Treble to mitigate this. Project Treble introduced a vendor interface that separates the core Android OS framework from the low-level, vendor-specific software. Because of this, an independent update of the operating system can be installed without the need for hardware vendors to re-certify drivers. This accelerated the deployment of security patches across diverse devices \cite{googleblogHereComes}.

Project Mainline further divides this modular design into smaller, updateable modules. The system can now update core components 
as standalone packages using the existing Google Play infrastructure. This approach shifts the security model from a large-scale update cycle to a continuous delivery system. While it improves security from emerging threats, it also introduces challenges for energy research. As these modules are standardized across the ecosystem, they may not be optimized for older or specialized hardware, which can cause discrepancies in energy efficiency after updates \cite{androidMainlineAndroid}.

\subsection{The Convergence of Security and Energy Constraints}
The evolution of Android’s architecture from the isolation provided by the Linux kernel to the modular updates of Project Mainline shows a shift toward stronger security. However, this increased complexity increased resource demands. The interaction between power-saving features 
and the continuous processing required for encryption and background security monitoring creates a trade-off. As Android seeks to balance data protection with battery capacity, it becomes essential to understand the overhead of these security mechanisms.



\section{Related Work} \label{sec:related}
The trade-off between mobile security and energy efficiency has been an active research area.

Initial studies showed that the energy consumption in security operations is disproportionately related to the computational complexity of cryptographic algorithms. Potlapally et al. \cite{potlapally_2006} have shown that asymmetric encryption consumes more energy than symmetric encryption, like AES, due to the higher number of CPU cycles required per bit of data. A recent study conducted by Ferreira et al. \cite{relwork1} showed a rapid increase in energy consumption with high-level security operations for data encryption on Android. 

Götzfried and Müller \cite{relwork5} worked on storage security and showed the difference between full disk encryption (FDE) and File-based encryption (FBE) in Android. They found that the FBE can be a reason for the decrease in battery capacity, as during FBE, the kernel's cryptography API is invoked very frequently during I/O operations in an encrypted system. 


Research reveals that memory usage, execution time, and CPU activity act as major contributors to power consumption on mobile applications. Various studies conducted across Android and iOS platforms observed a positive correlation between memory utilization and energy consumption, and execution time was found to influence battery usage depending on the type of application \cite{relwork2}. These findings suggest that runtime optimization and efficient resource utilization are essential factors in developing sustainable mobile applications. However, Google introduced an inline encryption hardware which mitigated some of this overhead by taking away some decryption tasks from the main CPU \cite{google2025fbe}. 

Nidawi et al. \cite{relwork3} conducted a review of energy consumption by applications in the Android platform. They showed that software architecture, application behavior, interface design, hardware resource usage, and network technologies are the main factors in mobile energy consumption. They pointed out that application design decisions play a crucial role in battery consumption and overall system efficiency.

Schweizer et al. \cite{relwork4} used sensors in mobile tracking systems to track the energy consumption in their work. They found that continuous sensor access increases power consumption and highlighted that computational workload and continuous resource access are the major reasons for energy consumption in mobile devices.

The review of various research works revealed that very limited research has been done to investigate the relationship between deep learning-based Android malware detection models and their runtime energy efficiency on the devices. We also found that existing malware detection approaches focus mainly on detection efficiency and overlook the deployment concerns. 


In this work, we evaluated the trade-off between malware detection effectiveness and energy efficiency using deep-learning models deployed on Android devices and found an equilibrium between detection effectiveness and energy efficiency.  


\section{Methodology} \label{sec:research_methodology}
\begin{figure*}[h]
    \centering
    \includegraphics[width=0.9\textwidth]{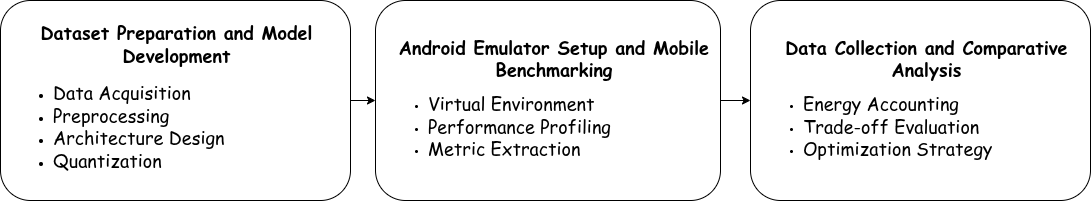}
    \caption{Methodology}
    \label{fig:methodology}
\end{figure*}

This section explains the experimental steps used to evaluate the tradeoff between detection efficiency and power consumption. We implemented multiple Android malware detection models and evaluated their classification performance along with runtime efficiency on an Android-based environment in this study. The methodology is shown in the Figure \ref{fig:methodology}. As shown in the figure, the methodology consists of three phases: dataset preparation and model development, mobile deployment and benchmarking, and performance analysis. Dataset preparation and model development are the foundational phases of the methodology, where datasets were collected and preprocessed, and Multi-Layer Perceptron (MLP) models were developed and optimized. The mobile deployment phase creates an environment where the models are tested for efficiency. It includes virtual environment creation, performance profiling, and metric extraction for further analysis. The performance analysis phase analyzes the metrics obtained from the second phase and identifies the optimum balance between energy consumption and detection accuracy. It consists of energy accounting, trade-off evaluation, and optimization strategy. The next subsections explain them in detail.

\subsection{Dataset Preparation and Model Development}
This is the first phase of this work. The purpose of this phase was to develop effective malware detection models and their optimization for Android devices. This process is divided into four steps.

Data acquisition and preprocessing are the first step. We used two publicly available 
datasets: TUANDROMD dataset \cite{tuandromd_dataset} and DREBIN dataset \cite{mlsecDrebinDataset}. TUANDROMD focuses on behavioural features and malicious activity patterns. The DREBIN dataset is created with 
features extracted from \verb|AndroidManifest.xml| and disassembled dex code (e.g., API calls, permissions, and intents). 

Although these datasets are already clean, we performed some data cleaning, such as handling null or missing values. We also encoded the categorical labels for malware and benign classes to binary labels of 1 and 0. After this step, the final TUANDROMD dataset contains 4465 samples and 242 attributes, and DREBIN dataset contains 7249 samples and 215 attributes.  

The next step was architectural design. We developed a lightweight Deep Learning (DL) architecture using Multi-Layer Perceptron (MLP) models for a classification task. 
We developed three variants of these models, consisting of 3, 4, and 5 layers, to evaluate the impact of model depth on the experiment. Each layer used the ReLU (Rectified Linear Unit) activation function, and the output layer used the Sigmoid function for binary classification. 

Model optimization was the next step. We applied Post-Training Quantization (PTQ) to convert the model weights from float32 to INT8 to reduce the model’s memory footprint and accelerate runtime on mobile CPUs.
We used 5-fold cross-validation to improve the model's reliability. Classification performance of the models was evaluated using accuracy, precision, recall, F1-score, and Matthews Correlation Coefficient (MCC).

This optimized pipeline made sure that the developed models not only perform better in terms of accuracy but also consume less power.




\subsection{Android Emulator Setup and Mobile Benchmarking}
This phase was devoted to evaluating the models developed in the previous phase on an Android device. This phase consists of three steps. In the first step, we created a controlled environment to deploy the models on Android Virtual Device (AVD). We simulated a Google Pixel 6 smartphone with an x86\_64 Google Play system image with 4 GB RAM and 25 GB internal storage for the benchmarking experiment.

The second step is the performance profiling. In this step, the classification models were converted into the TensorFlow Lite (TFLite) format, as this format is optimized for on-device execution. These \verb|.tflite| models were integrated into a custom test suite in the AI Benchmark application \cite{aibenchmarkAIBenchmark}.  All TFLite models were then pushed to the emulator and executed using the CPU backend. We tested full precision MLP models in FP32 mode, and quantized models in INT8 mode. 

This is the third and final step of this phase. It involves the extraction of metrics that define the efficiency of the model. We experimented with each model 10 times and recorded the average for each metric to minimize random variation and achieve reliable estimates of each metric. The extracted metrics include accuracy, precision, recall, F1-score, MCC, average latency, throughput, and model size.

\subsection{Data Collection and Comparative Analysis}
This is the third and final phase of our methodology. It focuses on combining the metrics collected in the previous phase to evaluate the sustainability of the model. This phase acts as a bridge between detection performance and energy efficiency. It consists of three steps. The first step is energy accounting. Since AVD environment doesn't have the ability for physical battery instrumentation, we used the approach given by Hu et al. \cite{HU2020681}. They proposed a formula to calculate energy. As per their work, energy \textit{E} required for a secure operation can be calculated as per the equation \ref{eq_energy_calculation}. 

Trade-off evaluation was the next step. Here, we performed a comparative analysis between classification metrics and energy-related metrics. This step helped us in understanding the trade-off between detection and energy consumption.

The last step in this phase was identifying the optimization strategy. In this step, we identified the configurations that provided a high enough detection performance with minimum energy consumption. This optimization step was crucial as it provided a configuration with which an effective detection model can be created with less overhead on devices.



\section{Results and Discussion} \label{sec:results_n_discussion}
This section shows the results of the conducted experiments. The results are categorized into malware detection performance and the consumption of energy during this process. The next subsection shows these results.

\subsection{Classification Performance Analysis}
The first evaluation is focused on the detection efficiency of the Multi-Layer Perceptron (MLP) and Quantized Neural Network (QNN) architectures using the TUANDROMD and DREBIN datasets. Table \ref{tab:classification_results} shows the performance at different layer depths using both datasets.

\begin{table*}[!t]
\vspace{3mm}
\centering
\caption{Classification Performance across Model Architectures and Datasets}
\label{tab:classification_results}
\begin{tabular}{@{}llcccccc@{}}
\toprule
\textbf{Dataset} & \textbf{Model Type} & \textbf{Layers} & \textbf{Acc. (\%)} & \textbf{Prec.} & \textbf{Rec.} & \textbf{F1} & \textbf{MCC} \\ \midrule
TUANDROMD & MLP (FP32) & 3 & 99.28 & 1.00 & 1.00 & 1.00 & 0.9777 \\
 & MLP (FP32) & 4 & 99.19 & 0.99 & 1.00 & 0.99 & 0.9748 \\
 & MLP (FP32) & 5 & \textbf{99.31} & 0.99 & 1.00 & 1.00 & \textbf{0.9784} \\
 & QNN (INT8) & 3 & 99.24 & 0.99 & 1.00 & 1.00 & 0.9763 \\
 & QNN (INT8) & 4 & 99.22 & 0.99 & 1.00 & 1.00 & 0.9755 \\
 & QNN (INT8) & 5 & \textbf{99.31} & 0.99 & 1.00 & 1.00 & \textbf{0.9784} \\ \midrule
DREBIN & MLP (FP32) & 3 & 98.48 & 0.99 & 0.99 & 0.99 & 0.9575 \\
 & MLP (FP32) & 4 & 98.62 & 0.99 & 0.99 & 0.99 & 0.9614 \\
 & MLP (FP32) & 5 & 98.41 & 0.99 & 0.99 & 0.99 & 0.9555 \\
 & QNN (INT8) & 3 & 98.55 & 0.99 & 0.99 & 0.99 & 0.9587 \\
 & QNN (INT8) & 4 & \textbf{98.63} & 0.99 & 0.99 & 0.99 & \textbf{0.9618} \\
 & QNN (INT8) & 5 & 98.51 & 0.99 & 0.99 & 0.99 & 0.9583 \\ \bottomrule
\end{tabular}
\end{table*}


As shown in the Table \ref{tab:classification_results}, both architectures performed well. On the TUANDROMD dataset, the highest accuracy achieved is 99.31\% with a 5-layer model, and the recall is constant 1.00. This shows that all malicious samples were detected successfully. The DREBIN dataset also showed a similar result with the highest accuracy of 98.63\% with a 4-layer quantized model. The change from FP32 to INT8 quantization caused negligible accuracy loss, and in some DREBIN cases, there was a marginal improvement. It suggests that quantized models are small but equally reliable for Android malware detection.

\subsection{Resource and Energy Consumption Analysis}
We extended our experiment from detection efficiency to the model's resource usage to address the trade-off between security and energy efficiency. Since AVD environment doesn't have the ability for physical battery instrumentation, we used the equation proposed by Hu et al. \cite{HU2020681} and explained in Equation \ref{eq_energy_calculation}. The results are shown in Table \ref{tab:energy_consumption_results}.

\begin{table*}[!t]
\centering
\caption{Computational Efficiency and Estimated Energy Overhead}
\label{tab:energy_consumption_results}
\begin{tabular}{@{}llccccc@{}}
\toprule
\textbf{Dataset} & \textbf{Model Type} & \textbf{Layers} & \textbf{Size (KB)} & \textbf{Throughput (inf/s)} & \textbf{$t_{enc}$ (ms)} & \textbf{Energy (mJ)} \\ \midrule
TUANDROMD & MLP (FP32) & 3 & 405.42 & 41,667 & 0.0240 & 0.0238 \\
 & MLP (FP32) & 4 & 413.82 & 45,455 & 0.0220 & 0.0219 \\
 & MLP (FP32) & 5 & 416.20 & 40,000 & 0.0250 & 0.0248 \\
 & QNN (INT8) & \textbf{3} & \textbf{115.46} & \textbf{47,619} & \textbf{0.0210} & \textbf{0.0189} \\
 & QNN (INT8) & 4 & 118.78 & 43,478 & 0.0230 & 0.0206 \\
 & QNN (INT8) & 5 & 120.17 & 38,462 & 0.0260 & 0.0231 \\ \midrule
DREBIN & MLP (FP32) & 3 & 379.42 & 45,455 & 0.0220 & 0.0219 \\
 & MLP (FP32) & 4 & 387.82 & 44,444 & 0.0225 & 0.0224 \\
 & MLP (FP32) & 5 & 390.20 & 41,667 & 0.0240 & 0.0238 \\
 & QNN (INT8) & 3 & 108.98 & 35,714 & 0.0280 & 0.0248 \\
 & QNN (INT8) & \textbf{4} & \textbf{112.28} & \textbf{38,462} & \textbf{0.0260} & \textbf{0.0231} \\
 & QNN (INT8) & 5 & 113.67 & 36,364 & 0.0275 & 0.0244 \\ \bottomrule
\end{tabular}
\end{table*}

As shown in the Table \ref{tab:energy_consumption_results}, the results show a trade-off between architectural complexity and energy sustainability. 

\subsubsection{Energy-Latency Correlation}
As per the data shown in the table, energy consumption ($E$) showed a linear relation with latency ($t_{enc}$). For the TUANDROMD dataset, when model depth was increased from 3 to 5 layers in the quantized setup, the $t_{enc}$ also got increased from 0.0210 ms to 0.0260 ms. This caused the increase of energy consumption from 0.0189 mJ to 0.0231 mJ which is an increase of 22\%. This shows that minimizing the time of CPU being in high power state is an effective way to reduce the energy usage for background security services on an Android device.

\subsubsection{Efficiency Gains through Quantization}
Computational efficiency increased when the model precision was changed from FP32 to INT8. Quantization reduces the size of model size by about 3.5 times in all test cases. For example, TUANDROMD model size was reduced from 405.42 KB to 115.46 KB. Due to this reduction in model size, RAM consumption was also reduced, which improved the processing speed. The 3-layer QNN achieved a maximum throughput of 47619 inferences per second, which is about 14\% faster than the FP32 version. This makes real-time malware scanning possible with very little effect on user experience.  

\subsubsection{Identification of the Optimized Condition}
Comparison of the performance results in Table \ref{tab:classification_results} with the efficiency result in Table \ref{tab:energy_consumption_results} gave the best model setup for each dataset.

\begin{itemize}
    \item The 3-layer QNN was the optimal choice for TUANDROMD dataset as it consumed less energy (0.0189 mJ) with an accuracy of 99.24\%.

    \item The 4-layer QNN provided the optimal balance between accuracy and efficiency, as it achieved 98.63\% accuracy with 0.0231 mJ energy consumption. 
\end{itemize}

These results show that shallow quantized neural networks can solve the issue of tradeoff in Android malware detection due to their capability of providing strong security with less energy usage of resource-constrained Android phones.

\section{Conclusion} \label{sec:conclusion}

This paper optimizes the trade-off between security and energy consumption by finding suitable configurations for deploying deep learning malware detection models on Android devices. The work in this paper is focused on balancing detection performance and resource efficiency using post-training quantization (PTQ) and model design adjustments for the models created with the TUANDROMD and DREBIN datasets.

The experimental results show that strong security performance may not require high energy use. In the experiment, detection accuracy reached upto 99.31\% by optimizing the model architecture. The analysis also displayed that INT8 quantization is responsible for sustainable security, as due to this, the model achieved a 3.5$\times$ reduction in size while reducing the energy usage to 0.0189 mJ.


Our analysis identified the optimal configuration for both the datasets, which is a 3-layer QNN for TUANDROMD and a 4-layer QNN for DREBIN. These configurations reduced the energy consumption of security-related operations by increasing the throughput and reducing the duration of the CPU being in a high-power state. This helped in maintaining the trade-off.

Overall, this work shows that an optimally efficient setup for mobile security can be achieved with the combination of lower bit-width operations and balanced network depth. Under these conditions, the malware scanner has minimal impact on battery life
while maintaining good detection performance. This provides practical guidance for developers to build efficient and reliable real-time protection systems for Android devices.


While this work identifies optimized configurations for on-device malware detection, there are limitations of this work that provide directions for future work. A key limitation is the use of Android Virtual Devices (AVD) for performance and energy evaluation, as emulators do not fully capture real-world hardware behavior.
To address this limitation, future work will evaluate the INT8 models on real smartphones.

We will also study the effect of using Neural Processing Units (NPUs) for running the model. We expect that these specialized chips will further reduce energy use during continuous background scanning, which will improve overall efficiency.
In addition, the impact of FP32-to-INT8 quantization on the calibration and reliability of predictive confidence in deep learning models for malware classification tasks will be investigated.


\balance
\bibliographystyle{plain} 
\bibliography{references}

@misc{statcounterMobileOperating,
	author = {},
	title = {{M}obile {O}perating {S}ystem {M}arket {S}hare {W}orldwide | {S}tatcounter {G}lobal {S}tats --- gs.statcounter.com},
	howpublished = {\url{https://gs.statcounter.com/os-market-share/mobile/worldwide}},
	year = {},
	note = {[Accessed 18-03-2026]},
}

@misc{S_R19,
	author = {Ignatov, A. and others},
	title = {{ICML 2026 Call For Papers}},
	howpublished = {International Conference on Machine Learning},
	year = {2026},
	url = {https://icml.cc/Conferences/2026/CallForPapers}
}

@article{S_R105,
	author = {Wang, S. and others},
	title = {{A Practical and Accurate Battery Emulator for Android Smartphones}},
	journal = {ResearchGate},
	year = {2025}
}

@article{S_R25,
	author = {Nguyen, T. T. and others},
	title = {{Time-and-Energy consumption offloading for mobile devices}},
	journal = {ResearchGate},
	year = {2024}
}

@inproceedings{S_R46,
	author = {Agarwal, M. and others},
	title = {{AI-Powered Android Malware Detection using Machine Learning}},
	booktitle = {Proceedings of the IEEE/CVF},
	year = {2025}
}

@misc{S_R20,
	author = {Seo, J. and others},
	title = {{AI for Sciences Track Call for Papers}},
	howpublished = {KDD 2026},
	year = {2026},
	url = {https://kdd2026.kdd.org/ai4sciences-track-call-for-papers/}
}

@misc{S_R61,
	author = {{GeeksforGeeks}},
	title = {{Beginner Friendly Machine Learning Projects 2025}},
	year = {2025},
	url = {https://www.geeksforgeeks.org/machine-learning/machine-learning-projects/}
}

@article{S_R195,
	author = {Anwar, H.},
	title = {{ARENA: A tool for measuring and analysing the energy efficiency of Android apps}},
	journal = {arXiv preprint arXiv:2510.01754},
	year = {2025}
}

@misc{S_R97,
	author = {Klieder, W. and others},
	title = {{Evaluating Static Analysis Alerts with LLMs}},
	howpublished = {CMU SEI Blog},
	year = {2024},
	url = {https://www.sei.cmu.edu/blog/evaluating-static-analysis-alerts-with-llms/}
}

@article{S_R178,
	author = {{IJIRSET}},
	title = {{Green AI Energy-Efficient Machine Learning Model Selector}},
	journal = {International Journal of Innovative Research in Science, Engineering and Technology},
	year = {2025}
}

@article{S_R102,
	author = {Mehrotra, D. and Srivastava, R. and Nagpal, R. and Nagpal, D.},
	title = {{Multiclass classification of mobile applications as per energy consumption}},
	journal = {Journal of King Saud University - Computer and Information Sciences},
	volume = {33},
	number = {6},
	pages = {719--727},
	year = {2021},
	doi = {10.1016/j.jksuci.2018.05.007}
}

@misc{S_R37,
	author = {{ICLR}},
	title = {{Societal Considerations of Representation Learning}},
	year = {2026},
	url = {https://iclr.cc/}
}

@article{S_R66,
	author = {Friha and others},
	title = {{Model Quantization for Edge Environments}},
	journal = {arXiv preprint arXiv:2504.03360},
	year = {2025}
}

@article{S_R173,
	author = {{CarbonTracker}},
	title = {{Standardized Approach for Evaluating AI Model Inference Efficiency}},
	journal = {MDPI Sensors},
	volume = {25},
	number = {3},
	year = {2025}
}

@techreport{S_R210,
	author = {{DTU}},
	title = {{A Survey of Quantization Techniques in Embedded AI Toolchains}},
	institution = {Technical University of Denmark},
	year = {2025}
}

@misc{androidPlatformArchitecture,
	author = {},
	title = {{P}latform architecture  |  {A}ndroid {D}evelopers --- developer.android.com},
	howpublished = {\url{https://developer.android.com/guide/platform}},
	year = {},
	note = {[Accessed 30-01-2026]},
}

@techreport{53044,
	title	= {The Android Platform Security Model (2023)},author	= {René Mayrhofer and Jeff Vander Stoep and Chad Brubaker and Dianne Hackborn and Bram Bonné and Güliz Seray Tuncay and Roger Piqueras Jover and Michael Specter},year	= {2023},URL	= {https://arxiv.org/pdf/1904.05572/1000.pdf},institution	= {Cornell University}}

@misc{androidPowerManagement,
	author = {},
	title = {{P}ower management  |  {A}ndroid {O}pen {S}ource {P}roject --- source.android.com},
	howpublished = {\url{https://source.android.com/docs/core/power/mgmt}},
	year = {},
	note = {[Accessed 30-01-2026]},
}

@article{HU2020681,
	title = {Taming energy cost of disk encryption software on data-intensive mobile devices},
	journal = {Future Generation Computer Systems},
	volume = {107},
	pages = {681-691},
	year = {2020},
	issn = {0167-739X},
	doi = {https://doi.org/10.1016/j.future.2017.09.025},
	url = {https://www.sciencedirect.com/science/article/pii/S0167739X17320113},
	author = {Yang Hu and John C.S. Lui and Wenjun Hu and Xiaobo Ma and Jianfeng Li and Xiao Liang}
}

@misc{googleblogHereComes,
	author = {},
	title = {{H}ere comes {T}reble: {A} modular base for {A}ndroid --- android-developers.googleblog.com},
	howpublished = {\url{https://android-developers.googleblog.com/2017/05/here-comes-treble-modular-base-for.html}},
	year = {},
	note = {[Accessed 30-01-2026]},
}

@misc{androidMainlineAndroid,
	author = {},
	title = {{M}ainline  |  {A}ndroid {O}pen {S}ource {P}roject --- source.android.com},
	howpublished = {\url{https://source.android.com/docs/core/ota/modular-system}},
	year = {},
	note = {[Accessed 30-01-2026]},
}

@misc{tuandromd_dataset,
	author = {},
	title = {TUANDROMD Dataset},
	howpublished = {\url{https://archive.ics.uci.edu/dataset/855/tuandromd+(tezpur+university+android+malware+dataset)}},
	year = {},
	note = {[Accessed 30-04-2026]},
}

@misc{mlsecDrebinDataset,
	author = {Daniel Arp},
	title = {{T}he {D}rebin {D}ataset},
	howpublished = {\url{https://drebin.mlsec.org/}},
	year = {},
	note = {[Accessed 30-04-2026]},
}

@misc{aibenchmarkAIBenchmark,
	author = {},
	title = { {A}{I}-{B}enchmark  --- ai-benchmark.com},
	howpublished = {\url{https://ai-benchmark.com/download}},
	year = {},
	note = {[Accessed 30-04-2026]},
}

@inproceedings{relwork1,
  author={Ferreira, Joao and Santos, Bernardo and Oliveira, Wellington and Antunes, Nuno and Cabral, Bruno and Fernandes, João Paulo},
  booktitle={2023 IEEE/ACM 10th International Conference on Mobile Software Engineering and Systems (MOBILESoft)}, 
  title={On Security and Energy Efficiency in Android Smartphones}, 
  year={2023},
  volume={},
  number={},
  pages={87-95},
  doi={10.1109/MOBILSoft59058.2023.00018}}

@inproceedings{relwork2,
author = {Rozikin, Khoirur and Santoso, Joseph and Migunani, Migunani and Hassan S, Noorul},
year = {2025},
month = {08},
pages = {1-7},
title = {Comparative Energy Efficiency Analysis of Mobile Apps on Android and iOS Platforms},
doi = {10.1109/ICCIT65724.2025.11166978}
}

@article{relwork3,
author = {al Nidawi, Hasan and Koh, Tieng Wei and Dawood, Kareem and Khaleel, Ammar},
year = {2017},
month = {12},
pages = {6776-6787},
title = {Energy consumption patterns of mobile applications in android platform: A systematic literature review},
volume = {95},
journal = {Journal of Theoretical and Applied Information Technology}
}

@inproceedings{relwork4,
  author={Schweizer, Immanuel and Bärtl, Roman and Schmidt, Benedikt and Kaup, Fabian and Mühlhäuser, Max},
  booktitle={6th International Conference on Mobile Computing, Applications and Services}, 
  title={Kraken.me mobile: The energy footprint of mobile tracking}, 
  year={2014},
  volume={},
  number={},
  pages={82-89},
  doi={10.4108/icst.mobicase.2014.257823}}

@article{relwork5,
author = {Götzfried, J. and Müller, T.},
year = {2014},
month = {03},
pages = {84-100},
title = {Analysing Android's full disk encryption feature},
volume = {5},
journal = {Journal of Wireless Mobile Networks, Ubiquitous Computing, and Dependable Applications}
}

@ARTICLE{potlapally_2006,
  author={Potlapally, N.R. and Ravi, S. and Raghunathan, A. and Jha, N.K.},
  journal={IEEE Transactions on Mobile Computing}, 
  title={A study of the energy consumption characteristics of cryptographic algorithms and security protocols}, 
  year={2006},
  volume={5},
  number={2},
  pages={128-143},
  doi={10.1109/TMC.2006.16}}

@online{google2025fbe,
	author    = {{Google}},
	title     = {File-based encryption},
	journal   = {Android Open Source Project (AOSP)},
	year      = {2025},
	url       = {https://source.android.com/docs/security/features/encryption/file-based},
	note      = {Accessed: 2026-01-30}
}

\end{document}